**Adjusting confidence intervals under covariate-adaptive randomization in non-inferiority and equivalence trials**

Masahiro Kojima, Ph.D.[1]

Hirotaka Mano, MEng[1]

Kana Yamada, MEng[1]

Keisuke Hanada, MSc[1]

Yuji Tanaka, MEng[1]

Junji Moriya, Ph.D.[1]

[1]Biometrics Department, R&D Division, Kyowa Kirin Co., Ltd., Tokyo, Japan.

**Corresponding author**

Name: Masahiro Kojima, Ph.D.

Address: Biometrics Department, R&D Division, Kyowa Kirin Co., Ltd.

Otemachi Financial City Grand Cube, 1-9-2 Otemachi, Chiyoda-ku, Tokyo, 100-004, Japan.

Tel: +81-3-5205-7200

FAX: +81- 3-5205-7182

Email: masahiro.kojima.tk@kyowakirin.com

**Running title**: Adjusting confidence intervals under covariate-adaptive randomization

**Financial support**: None

**A conflict of interest disclosure statement**: Authors are employees of Kyowa Kirin Co., Ltd.

**Word count**: 5263

**The number of figures**: 1

**The number of tables**: 3

**Declarations:**

Consent for publication: All authors have agreed.

Availability of data and materials: Not applicable.

Funding: Not applicable.

Authors' contributions: M. Kojima: Conception and design; development of methodology; writing, review, and revision of the manuscript; administrative, technical; and study supervision. H. Mano: Analysis and interpretation of data (e.g., statistical analysis, biostatistics, computational analysis), writing, review, and revision of the manuscript. K. Yamada: Analysis and interpretation of data (e.g., statistical analysis, biostatistics, computational analysis), writing, review, and revision of the manuscript. K. Hanada: Development of methodology analysis and interpretation of data (e.g., statistical analysis, biostatistics, computational analysis), writing, review, and revision of the manuscript. Y. Tanaka: Analysis and interpretation of data (e.g., statistical analysis, biostatistics, computational analysis), and review. J. Moriya: Writing, review and study supervision.

Patients are directly involved in this study: Not applicable.

**Abstract**

Regulatory authorities guide the use of permutation tests or randomization tests so as not to increase the type-I error rate when applying covariate-adaptive randomization in randomized clinical trials. For non-inferiority and equivalence trials, this paper derives adjusted confidence intervals using permutation and randomization methods, thus controlling the type-I error to be much closer to the pre-specified nominal significance level. We consider three variable types for the outcome of interest, namely normal, binary, and time-to-event variables for the adjusted confidence intervals. For normal variables, we show that the type-I error for the adjusted confidence interval holds the nominal significance level. However, we highlight a unique theoretical challenge for non-inferiority and equivalence trials: binary and time-to-event variables may not hold the nominal significance level when the model parameters are estimated by models that diverge from the data-generating model under the null hypothesis. To clarify these features, we present simulation results and evaluate the performance of the adjusted confidence intervals.

# 1. Introduction

Randomization is the gold standard for evaluating the treatment effect in clinical trials, because it statistically reduces any potential biases between groups. Although complete randomization, in which all subjects are assigned to a group under equal probability, is easy to achieve, the disadvantage is that the number of subjects cannot be balanced between treatment groups[1]. Stratified randomization categorizes subjects into strata according to covariate values prior to randomization, and then assigns subjects to a treatment within each stratum. The aim of stratified randomization is to improve the efficiency of estimators and the power of statistical tests in small trials[1]. However, the power tends to decrease when there is a stratification factor with a small number of subjects or when the effect of the stratified factor is large[2]. Covariate-adaptive randomization (CAR) is frequently used in clinical trials as it balances the number of subjects between treatment groups with covariate information. Taves[3] proposed the minimization procedure, which is one of the oldest CAR methods[3]. Pocock and Simon[5] generalized the dynamic minimization procedure using the allocation probability for treatment groups, an approach that was applied in over 400 clinical trials from 1989 to 2008[4]. These CARs have been used in clinical trials for superiority comparisons as well as for non-inferiority and equivalence comparisons[18,19,20,21]. Because of its versatility,

CAR is applicable in many clinical situations and is "particularly useful when (1) many baseline factors are known to have a major impact on the patient's prognosis, (2) many centers accrue patients into the trial, and the centers may contribute very different numbers of patients, or (3) the trial sample size is small"[15].

Any randomization method should be associated with valid statistical tests under the randomization scheme. A test is said to be valid if the type-I error is controlled within a specific significance level and conservative if the type-I error is too small compared with the specific significance level. Although test procedures can generally be valid under complete randomization, this is not guaranteed under other randomization schemes. For example, the two-sample t-test is conservative but not valid under CAR[6] because a balanced allocation of CAR makes the test statistic appear closer to the value under the null hypothesis[7]. Thus, in clinical trials with CAR, the test procedure under the model associated with the applied randomization should be used. However, in practice, the analysis model may not include all allocation factors, such as clinical sites. Hence, an alternative method is required.

The Food and Drug Administration (FDA) and the Center for Drug Evaluation (CDE) have published guidelines for the testing procedure when using CAR. The FDA's guidelines on "Adaptive Design Clinical Trials for Drugs and Biologics Guidance for Industry" state that "Covariate-adaptive treatment assignment techniques do not directly increase the Type I error probability when analyzed with the appropriate methodologies (generally randomization or permutation tests)"[22], while the CDE's "Guidelines for Randomized Assignment of Drug Clinical Trials (for Trial Implementation)" note that "Attention should also be paid to the use of appropriate statistical analysis methods (e.g., random or permutation tests) and to avoid increasing the Type I error rate"[23]. These two regulatory agencies have suggested using randomization tests or permutation tests in trials using CAR.

Permutation and randomization tests are simulation-based methods that approximately identify rejection points better than a mathematical distribution. Permutation tests use an empirical distribution of test statistics derived from multiple randomly permutated datasets to test the null hypothesis. Randomization tests are similar, except that the empirical distribution of test statistics is derived by multiple re-randomized datasets according to a pre-specified randomization scheme. Permutation and randomization tests are often mistakenly believed to be equivalent; however, under non-complete randomization, permutation tests obtain the empirical distribution by completely reordering the respective data at random, whereas randomization tests use re-randomization to obtain the empirical distribution[8, 9].

These test procedures are a valid means of demonstrating the superiority of a treatment effect between treatment groups[10], but to the best of our knowledge no reports have evaluated the validity of the test procedures in non-inferiority and equivalence trials. Permutation and randomization tests simulate the distribution of test statistics with no treatment effect between treatment groups. However, in testing for non-inferiority and equivalence, it may be inappropriate to apply the simulated distribution because the null hypothesis for non-inferiority and equivalence tests relates to a nonzero treatment effect between treatment groups.

From the issues identified above regarding CAR, permutation tests, and randomization tests, we consider the following three research questions:

(i) Are non-inferiority and equivalence tests conservative under CAR?

(ii) Are non-inferiority and equivalence tests with a permutation test or randomization test valid under CAR?

(iii) How are the confidence intervals for the permutation method or randomization method derived?

To address these research questions, we prove the validity and conservativeness of the test procedures for (i) and (ii) by referring to Shao (2010)[6], and introduce the estimation of permutation test- and randomization test-based confidence intervals for (iii). The remainder of this paper is organized as follows. Section 2 introduces the relevant randomization methods and test procedures for non-inferiority and equivalence tests. Section 3 describes the formulation of computer simulations for evaluating the performance of our proposed adjusted confidence intervals for non-inferiority and equivalence tests, before Section 4 presents the simulation results. Finally, Section 5 summarizes the conclusions to this study and discusses our findings.

## 2. Methods

Initially, we revisit the procedure proposed by Pocock and Simon[5] and the stratified permuted block randomization (SPBR) technique. We define the following notation to describe the randomization methods: $t = 1$ denotes the treatment group and $t = 0$ denotes the control group; the randomization factors involve $m$ discrete covariates $\mathbf{Z} = (Z_1, Z_2, \ldots, Z_m)$; and the observed discrete covariates for the $i$-th subject are given by $\mathbf{z}_{(i)}$.

For Pocock and Simon's procedure, we assume that the randomized assignment for the first $(n-1)$ subjects has been completed. To determine the group for the $n$-th subject, we calculate the cumulative value for the observed discrete covariate of the $n$-th subject in each treatment group as $D(n) = \sum_{i=1}^{m} f(N_{z_{(i)}1}(n) - N_{z_{(i)}0}(n))$, where $N_{z_{(i)}t}$ is the cumulative number of subjects up to the $n$-th subject corresponding to the observed covariate $\mathbf{z}_{(i)}$ of the $n$-th subject for each treatment group, and the function $f(x)$ is defined as $f(x) = x$ for Taves's minimization method[3] or $f(x) = |x|$ for Pocock and Simon's procedure. To modify the probability $p$ of the $n$-th subject being assigned to $t = 1$, we use a technique similar to Efron's biased coin method. The specific probability is as follows:

1. If $D(n) = 0$, then $p = 0.5$
2. If $D(n) < 0$, then $p = q$, where $q \in (0.5, 1]$

3. If $D(n) > 0$, then $p = 1 - q$, where $q \in (0.5, 1]$

For Taves's minimization method, set $q = 1$.

For the SPBR technique, we assume that there are $c_j$ categories for $Z_j$ ($j = 1, 2, \ldots m$). Consequently, there are a total of $\prod_{j=1}^{m} c_j$ strata. Within each stratum, we determine the block size in advance and establish the number of groups required based on a certain proportion within the block. The order of the treatment group and control group within a block is randomly permutated among different blocks. The $i$-th subject is assigned to the group for the block corresponding to the stratum of $c_i$.

We evaluate non-inferiority using the following null and alternative hypotheses:

$$H_{NI,0}: \mu_1 - \mu_0 \leq -\delta_{NI} \text{ vs } H_{NI,1}: \mu_1 - \mu_0 > -\delta_{NI}, \text{ for } \delta_{NI} > 0$$

where $\mu_1$ is a true mean of the treatment group, $\mu_0$ is the true mean of the control group, and $\delta_{NI}$ denotes a pre-specified non-inferiority margin. We define the test statistic as $T(-\delta_{NI})$, which includes the non-inferiority margin. For example, when assuming that the outcomes are normally distributed, the test statistic is given by $T(-\delta_{NI}) = \frac{\overline{Y}_1 - \overline{Y}_0 + \delta_{NI}}{\sqrt{S^2\left(\frac{1}{n_1} + \frac{1}{n_0}\right)}}$, where $\overline{Y}_t$ and $n_t$ are the sample mean and sample size, respectively, for $t = 0, 1$, and $S^2$ is the pooled variance. Additionally, we assume that the outcomes are generated from a model in which all covariates are randomization factors. Shao et al.[6] suggested that the test statistic $T(-\delta_{NI})$ is conservative because the estimation of variance is unrelated to the non-inferiority margin. The asymptotic conservativeness of $T(-\delta_{NI})$ can be expressed as:

$$\lim_{n \to \infty} P\big(T(-\delta_{NI}) > c_\alpha | H_{NI,0}, \text{covariate adaptive}\big) < \alpha,$$

where $c_\alpha$ is the critical value for rejecting the null hypothesis and $\alpha$ is the significance level. Moreover, $T(-\delta_{NI})$ is valid in cases where the outcomes are adjusted by the model in which all covariates are randomization factors. For time-to-event (TTE) outcomes, $T(-\delta_{NI})$, as derived from Cox's regression model in which all covariates are randomization factors, is valid when the outcomes are generated by the model in which all covariates are randomization factors.

The null and alternative hypotheses for the equivalence test are:

$$H_{EQ,0}: \mu_1 - \mu_0 \leq -\delta_{EQ,L} \text{ or } \mu_1 - \mu_0 \geq \delta_{EQ,U} \text{ vs } H_{EQ,1}: -\delta_{EQ,L} < \mu_1 - \mu_0 < \delta_{EQ,U},$$

where $\mu_1$ is the true mean of the treatment group, $\mu_0$ is the true mean of the control group, and $\delta_{EQ,L}$, $\delta_{EQ,U}$ are positive equivalence margins. For normal variables, we assume that the outcomes are generated from a model in which all covariates are randomization factors. Using the same test statistics as for non-inferiority trials, $T(-\delta_{EQ,L})$ and $T(\delta_{EQ,U})$ are asymptotically conservative such that:

$$\lim_{n \to \infty} P\big(T(-\delta_{EQ,L}) > c_\alpha, T(\delta_{EQ,U}) < -c_\alpha | H_{EQ,0}, \text{covariate adaptive}\big) < \alpha,$$

because the pooled sample variance in CAR is larger than the variance in complete randomization[6]. Adjusting the generating model ensures that the test statistic is valid. For TTE variables, adjusting the generating model yields a valid test statistic.

Unlike superiority tests, which usually assume no difference between groups in the null hypothesis, non-inferiority and equivalence tests assume a specific difference between the groups in the null hypothesis. For example, when the model is mis-specified in cases of the logistic regression model for a binary outcome and the Cox regression model for a TTE outcome, the consistency of the parameter estimation with the true parameter value is lost[12,13]. If the true value of the parameter for the treatment group is exactly 0, i.e., there is no association between the outcome and the treatment group, no inconsistency would arise; however, such a situation would be unrealistic. Therefore, it becomes challenging (or difficult) to determine whether the type-I error can be controlled under the nominal significance level when the model for binary and TTE variables is misidentified. However, as mentioned in Section 1, our research has not found any reports that have evaluated this issue. Thus, it is meaningful to confirm this point through simulation studies.

Recall the FDA's requirement that "Covariate-adaptive treatment assignment techniques do not directly increase the Type I error probability when analyzed with the appropriate methodologies (generally randomization or permutation tests)" and the CDE's statement that "Attention should also be paid to the use of appropriate statistical analysis methods (e.g., random or permutation tests) and to avoid increasing the Type I error rate." To satisfy these requirements, we propose a method for deriving adjusted confidence intervals using the permutation method or randomization method. Our proposed method can be applied when the model is correctly identified or when the model is misidentified but still consistent for parameter estimation, such as in a linear regression model where the error term follows a normal distribution. To obtain adjusted confidence intervals that maintain the nominal significance level, we use empirical distributions for the permutation and randomization tests. Under CAR, the distribution of the test statistic is considered to deviate from the approximate distribution. Thus, we propose a method that constructs the $100(1-\alpha)\%$ confidence interval using the critical values of the upper $100\frac{\alpha}{2}\%$-th statistic and the lower $100\frac{\alpha}{2}\%$-th statistic in multiple statistics derived from multiple re-randomized or permutated datasets.

    The permutation method offers the exact distribution of the test statistics based on the independent design of the randomization method. We assume that the empirical distribution $F_p$ has an upper $100\alpha\%$ quantile point given by $q_p^\alpha$ for test statistics

generated from $\binom{n}{n/2}$ randomly permutated datasets. The type-I error for non-inferiority trials is expressed as:

$$\lim_{n \to \infty} P_{F_p}(T(-\delta_{NI}) > q_p^\alpha | H_{NI,0}, \text{covariate adaptive}) = \alpha.$$

For equivalence trials, the type-I error is

$$\lim_{n \to \infty} P_{F_p}(T(-\delta_{EQ,L}) > q_p^\alpha \text{ or } T(\delta_{EQ,U}) < q_p^{1-\alpha} | H_{EQ,0}, \text{covariate adaptive}) = \alpha.$$

The $100(1 - \alpha)\%$ confidence interval is derived using $q_p^\alpha$ and $q_p^{1-\alpha}$ instead of $c_\alpha$. For instance, in the case of a normally distributed variable, the adjusted $100(1 - \alpha)\%$ confidence interval is $\left[\bar{Y}_1 - \bar{Y}_0 - q_p^{1-\alpha}\sqrt{S^2\left(\frac{1}{n_1} + \frac{1}{n_0}\right)}, \bar{Y}_1 - \bar{Y}_0 + q_p^\alpha\sqrt{S^2\left(\frac{1}{n_1} + \frac{1}{n_0}\right)}\right]$. When $n$ is large, the number of permutated datasets should be sufficiently large to perform the analysis. The randomization method reassigns randomly permutated patients to treatment groups according to a predetermined randomization rule. We need to prepare a large number of randomly permutated observed datasets to obtain $F_r$. The upper $100\alpha\%$ quantile is $q_r^\alpha$. Similar to $F_p$, the empirical distribution $F_r$ ensures that the type-I error is within the nominal significance level using the critical values $q_r^\alpha$ and $q_r^{1-\alpha}$. The confidence intervals are obtained in the same way as for the permutation method.

We now present an intuitive interpretation about the distributions followed by the test statistic. Under CAR, the conservative test means that the distribution of the test statistic has a smaller variance than the normal distribution (i.e., the green line in エラー! 参照元が見つかりません。). Therefore, using ±1.96 on the red line as the rejection points, the type-I error is less than 5% because the confidence interval is too wide. Here, the randomization method can be used to obtain the approximate distribution of the test statistic and reproduce the distribution of the green line, thus controlling the type-I error at the nominal significance level. A similar point was made by Horiguchi and Uno[17] regarding the restricted mean survival time (RMST). For complete randomization, the RMST test is known to give an inflated type-I error under small samples[17]. This is because the distribution of the test statistic is heavy-tailed (like the blue line in Figure 1), although ±1.96 on the red line gives the rejection points. Because the distribution of the statistic by the permutation and randomization methods can be approximated by the blue line, using the blue line rejection points controls the type-I error.

## 3. Numerical simulation study

We conducted a simulation study to evaluate the performance of our proposed adjusted confidence interval in non-inferiority and equivalence tests. We consider three independent randomization factors. The variable $X_1$ is set to 1 for the new treatment group and to 0 for the standard treatment group; $X_2$ ranges from 1 to 10 corresponding to the value of $Z_1$; $X_3$ is the baseline continuous efficacy variable, distributed as $\mathcal{N}(30,5^2)$; and $X_4$ is 1 for male and 0 for female. The three variables $Z_1$, $Z_2$, and $Z_3$ were set as follows:

- $Z_1$ refers to the clinical site, with values assigned uniformly from 1 to 10.

- $Z_2$ refers to the disease status: "low" for $X_3 < 25$, "medium" for $25 \leq X_3 \leq 35$, and "high" for $X_3 > 35$.

- $Z_3$ refers to the patient's sex: male with a probability of 0.7 and female with a probability of 0.3.

CAR was conducted using Pocock and Simon's procedure. The assignment probability was set to $p = 0.7$ when there was a bias in the covariate between the two groups. SPBR was conducted using a block size of 4. The one-sided significance level was 2.5% and the two-sided significance level was 5.0%.

The sample size per group was 100. The number of simulations was 1,000. For each simulation, the permutation method and randomization method produced 1,000 data points. We reduced the number of simulations to limit the computational cost of the permutation and randomization methods. In addition, the simulation data used to evaluate non-inferiority and equivalence were kept the same for the purpose of efficient simulations. We now introduce the data-generating models and hypotheses for normal, binary, and TTE variables.

**Normal variables**

The data-generating model was:

$$Y = AX_1 + 2(X_2 - 5) + X_3 + 5X_4 + \epsilon, \epsilon \sim \mathcal{N}(0, 5^2),$$

where $A$ is set to $-3$ when evaluating type-I errors and $0$ when evaluating statistical power. We assumed that $\beta_1$ is a coefficient of the indicator variable $X_1$ for the treatment group tested in the statistical hypothesis.

The hypotheses for non-inferiority were:

$$H_{NI,0}: \beta_1 \leq -3 \text{ vs } H_{NI,1}: \beta_1 > -3.$$

The hypotheses for equivalence were:

$$H_{EQ,0}: \beta_1 \leq -3 \text{ or } \beta_1 \geq 3 \text{ vs } H_{EQ,1}: -3 < \beta_1 < 3.$$

We prepared three models to investigate the type-I error rates and statistical powers through the simulation study. The first was a naïve model without covariates that includes only the intercept and treatment group. The second model included clinically meaningful covariates used for practical situations (model 1). The third model was similar to the data-generating model (model 2).

[Naïve] $Y = \beta_0 + \beta_1 X_1$.

[Model 1] $Y = \beta_0 + \beta_1 X_1 + \beta_3 X_3 + \beta_4 X_4$, where $X_2$ is not included because covariates that classify multiple categories such as clinical sites are often excluded from the analysis model in practical situations.

[Model 2] $Y = \beta_0 + \beta_1 X_1 + \sum_{i=1}^{9} \beta_{2i} R_{2i} + \sum_{i=1}^{2} \beta_{3i} R_{3i} + \beta_4 X_4$, where the indicator variable $R_{2i}$ is 1 if the subject was enrolled at the $i$-th clinical site and 0 otherwise, the indicator variable $R_{31}$ is 1 if $R_3$ is low and 0 otherwise, and $R_{32}$ is 1 if $R_3$ is medium and 0 otherwise. This model was constructed using the concept that it should be close to the data generation model, but as simple as possible to avoid practically unrealistic situations. $\beta_1$, the naïve model, and models 1 and 2 were considered in the following settings for the same purpose as above.

**Binary variables**

The data-generating model was:

$$Y \sim Bin(n, p),$$

$$p = \frac{\exp(AX_1 + 0.2(X_2 - 5) + 0.05X_3 + 0.05X_4)}{1 + \exp(AX_1 + 0.2(X_2 - 5) + 0.05X_3 + 0.05X_4)},$$

where $A$ is set to $-1$ when evaluating type-I errors and $0$ when evaluating statistical power.

The hypotheses for the non-inferiority of the coefficient $\beta_1$ of $X_1$ were:

$$H_{NI,0}: \beta_1 \leq -1 \text{ vs } H_{NI,1}: \beta_1 > -1.$$

The hypotheses for equivalence were:

$$H_{EQ,0}: \beta_1 \leq -1 \text{ or } \beta_1 \geq 1 \text{ vs } H_{EQ,1}: -1 < \beta_1 < 1.$$

We prepared three logistic regression models to investigate the type-I error rates and statistical powers with Firth correction for the monotone likelihood problem[16].

[Naïve] $\log \frac{p}{1-p} = \beta_0 + \beta_1 X_1.$

[Model 1] $\log \frac{p}{1-p} = \beta_0 + \beta_1 X_1 + \beta_3 X_3 + \beta_4 X_4.$

[Model 2] $\log \frac{p}{1-p} = \beta_0 + \beta_1 X_1 + \sum_{i=1}^{9} \beta_{2i} R_{2i} + \sum_{i=1}^{2} \beta_{3i} R_{3i} + \beta_4 X_4.$

**Time-to-event variables**

The data-generating model was:

$$Y = \exp(AX_1 + 0.2(X_2 - 5) + 0.1X_3 + 0.5X_4 + \epsilon), \epsilon \sim \mathcal{N}(0,1),$$

where $A$ is set to $-0.5$ when evaluating type-I errors and $0$ when evaluating statistical power.

The censored data were generated such that $10\%$ were completely random. The cut-off time was set to $100$ because more than 90% of the events were observed.

The hypotheses for the non-inferiority of the coefficient $\beta_1$ of $X_1$ were:

$$H_{NI,0}: \beta_1 \leq -0.5 \text{ vs } H_{NI,1}: \beta_1 > -0.5.$$

The hypotheses for equivalence were:

$$H_{EQ,0}: \beta_1 \leq -0.5 \text{ or } \beta_1 \geq 0.5 \text{ vs } H_{EQ,1}: -0.5 < \beta_1 < 0.5.$$

We prepared Cox's regression model and an RMST model. For Cox's regression model:

[Naïve] $Y = h_0(t) \exp(\beta_1 X_1)$.

[Model 1] $Y = h_0(t) \exp(\beta_1 X_1 + \beta_3 X_3 + \beta_4 X_4)$.

[Model 2] $Y = h_0(t) \exp(\beta_1 X_1 + \sum_{i=1}^{9} \beta_{2i} R_{2i} + \sum_{i=1}^{2} \beta_{3i} R_{3i} + \beta_4 X_4)$.

$h_0(t)$ is a baseline hazard function.

For the RMST model, we applied Tian's covariate-adjustment method[14]. The estimator for the parameters of the RMST model for the data given by the generating model may not converge to the true value of $\beta_1$[14]. Therefore, we assumed that the true parameter value for the treatment group of the RMST model was $\bar{\beta}_1$, and calculated the approximate true value of the parameter $\bar{\beta}_1$ to be $-0.29122$ using 1,000,000 data points for each group. The restricted time was set to 80 because approximately 90% of events were observed.

The hypotheses for the non-inferiority of the coefficient $\bar{\beta}_1$ of $X_1$ were:

$$H_{NI,0}: \bar{\beta}_1 \leq -0.29122 \text{ vs } H_{NI,1}: \bar{\beta}_1 > -0.29122.$$

The hypotheses for equivalence were:

$$H_{EQ,0}: \bar{\beta}_1 \leq -0.29122 \text{ or } \bar{\beta}_1 \geq 0.29122 \text{ vs } H_{EQ,1}: -0.29122 < \bar{\beta}_1 < 0.29122.$$

[Naïve] $Y = \exp(\beta_0 + \bar{\beta}_1 X_1)$.

[Model 1] $Y = \exp(\beta_0 + \bar{\beta}_1 X_1 + \beta_3 X_3 + \beta_4 X_4)$.

[Model 2] $Y = \exp(\beta_0 + \bar{\beta}_1 X_1 + \sum_{i=1}^{9} \beta_{2i} R_{2i} + \sum_{i=1}^{2} \beta_{3i} R_{3i} + \beta_4 X_4)$.

We did not apply the log-rank test because it cannot adjust continuous covariates.

We evaluated the following three criteria to evaluate the results of the simulation scenarios explained above.

1. Average bias of the estimator of the parameter for the treatment group

2. Type-I error rate

3. Statistical power

## 4. Results

**Bias of estimator of parameter for treatment group.** The average bias results are presented in エラー! 参照元が見つかりません。. For normal variables, the absolute bias is less than 0.05. For binomial variables and Cox's model, the biases become smaller as the model moves closer to the data-generating model when there is a group difference. The biases are small when there is no difference and consistency of parameter estimation. For the RMST model, the biases are consistently small.

**Type-I error rate.** The type-I error rates are listed in エラー! 参照元が見つかりません。. For non-inferiority and equivalence, the simulation results are consistent because the same random number seed was used in the simulations. For normal variables, the type-I error rates are conservative in the naïve model and model 1, and slightly inflated in model 2 for the simple t-test under CAR and SPBR. The type-I error rates are adjusted when the randomization method is applied. However, the permutation method does not change the type-I error rates. For binomial variables and Cox's model, the type-I error rates are significantly inflated for the naïve model and model 1 because of the bias. Model 2 controls the type-I error rate well. For the RMST model, the type-I error rates are well-controlled overall, although there are a few cases of slight inflation.

**Statistical power.** The results for the statistical power are presented in エラー! 参照元が見つかりません。. Because the trend of the change in power is the same for non-inferiority and equivalence, we describe our results without distinction. For normal variables, the power increases as the model approaches the data-generating model. When the randomization method is used, the power is high for the naïve model and model 1. For binomial variables and Cox's model, the power decreases slightly as the model approaches the data-generating model. In addition, applying the randomization method and permutation method does little to change the power. For the RMST model, the power increases as the model approaches the data-generating model. When the randomization method is used, the power is slightly high for the naïve model and model 1.

## 5. Discussion

We have introduced adjusted confidence intervals using permutation and randomization methods for non-inferiority and equivalence trials. To obtain adjusted confidence intervals that maintain the nominal significance level, empirical distributions from the permutation and randomization methods were used. Under CAR, the distribution of the test statistic is considered to deviate from the approximate distribution, so we developed a method that constructs the $100(1-\alpha)\%$ confidence interval using the critical values of the upper $100\frac{\alpha}{2}\%$-th statistic and the lower $100\frac{\alpha}{2}\%$-th statistic in multiple statistics derived from multiple re-randomized or permutated datasets. However, there is some inconsistency in the estimation under model misspecification in the logistic and Cox regression models, and so adjusting for the critical values does not control for the type-I error rate.

Using these methods, we conducted simulations to evaluate whether permutation and randomization methods are appropriate for non-inferiority and equivalence trials under CAR. We considered normal, binary, and TTE variables (in Cox's and RMST models) as simulation scenarios under both the presence and absence of a treatment effect, and used a naïve model (only intercept and treatment group) as well as practical and close to data-generating models to investigate the simulation criteria. Under the assumption of no treatment effect, the bias was small for all variables and models. This is because there is no bias in the estimation if the true value of the coefficient parameter for the treatment group is 0. Under the assumption of a treatment effect (null hypothesis of non-inferiority or equivalence), the bias was also small in all analysis models for the normal and TTE (RMST) variables, and for the binomial and TTE (Cox) variables with model 2. A slightly larger bias was observed (i.e., ≥0.05) when the naïve model and model 1 were used for the binomial and TTE (Cox) variables. Randomization ensures that treatment effect estimates are unbiased even when covariates are omitted from a linear model, and our simulation results replicate this. The reason for the smaller bias in TTE (RMST) variables is that they maintain consistency even when the model is misidentified. For the binomial and TTE (Cox) variables, there have been reports of bias in cases where the covariates are omitted in logistic or Cox regression models under the presence of a treatment effect; the same results were observed in our simulations[12, 13]. In other words, in situations where there is an presence of a treatment effect, bias is likely to be introduced because the naïve model and model 1 with logistic or Cox regression do not estimate the parameters well.

For normal variables, although slightly inflated values of $T_{CAR}$ and $T_{SPBR}$ were obtained with model 2, the type-I error rates were adjusted under all other models and methods. The naïve model and model 1 were found to be quite conservative in terms of $T_{CAR}$, $PT_{CAR}$, $T_{SPBR}$, and $PT_{SPBR}$ because the distributions of the test statistic under these methods have lighter tails (i.e., the green line in Figure 1) and the 95% confidence intervals using ±1.96 standard deviations from the mean would be too wide. $RT_{CAR}$ and $RT_{SPBR}$ can avoid deflation of the type-I error, providing reasonable control by reproducing the distribution and shifting the rejection points closer to the nominal rejection value. Furthermore, the powers under these methods were higher than with the other methods. Using model 2 (i.e., close to the data-generating model), the type-I error rates were adjusted under almost all methods and the powers were relatively high.

The $T_{CAR}$ and $T_{SPBR}$ values with the binomial and TTE (Cox) variables indicate that the type-I error rates were controlled to be under or near the nominal level and the power was close to 80% in terms of non-inferiority, but relatively low in equivalence, when using model 2. However, the type-I error rates were not adjusted by the naïve model or model 1 because of the bias in the estimation of treatment effects. Similar trends were observed for the permutation method and randomization method. As for the TTE (Cox) variables and model 2, which could control the type-I error rate, the powers under $T_{CAR}$ and $T_{SPBR}$ were higher than under $PT_{CAR}$, $RT_{CAR}$, $PT_{SPBR}$, and $RT_{SPBR}$. Under the assumption of no treatment effect, Cox regression without any treatment using CAR would have resulted in a higher power because of type-I rate inflation. However, using permutation and randomization methods, the confidence intervals were closer to the nominal confidence interval, and the type-I error rates became valid and the powers were reduced accordingly. Sensitivity analyses may not be necessary under permutation and randomized methods as long as the model is close to the data-generating model. In terms of the binomial variable, there were no difference between the methods using model 2.

For the TTE (RMST) variables, the type-I error rates were controlled by the naïve model under $T_{CAR}$ and $T_{SPBR}$, but not under the other scenarios. According to Horiguchi and Uno[24], "there is notable inflation of the type I error rate with those asymptotic approaches when the sample size is small". Hence, the type-I error rates were not controlled under almost all scenarios because there were only 200 patients in this simulation. In addition, the type-I error rate under complete randomization was controlled by all three models (3.0% for naïve model, 3.9% for model 1, and 3.6% for model 2; these values are higher than under $T_{CAR}$, $PT_{CAR}$, $T_{SPBR}$, and $PT_{SPBR}$). This indicates that the distribution of the test statistic under complete randomization has a heavy tail (i.e., the blue line in Figure 1) and the 95% confidence interval using ±1.96 standard deviations would be too narrow. Horiguchi and Uno[24] reported a similar trend and found that implementing the permutation method could control the type-I error to some extent. We calculated the approximate true value of $\bar{\beta}_1$ using 1,000,000 data points for each group and used this as the criterion instead of $\beta_1$. In the actual situation, calculating the approximate true value is recommended because there is a difference between $\beta_1$ and $\bar{\beta}_1$, and the type-I error might not be controlled if we use $\beta_1$. When we calculate the number of subjects in a clinical trial using RMST, it is appropriate to use the approximate true value of the parameter to avoid inflation of the type-I error. The naïve model under $T_{CAR}$, $PT_{CAR}$, $T_{SPBR}$, and $PT_{SPBR}$ should be used under the condition of controlling the type-I error because these approaches have similar powers.

Regarding the difference between non-inferiority and equivalence, the power of equivalence is less than that of non-inferiority because testing with the upper confidence limit makes significance less likely. However, the type-I error rates were almost the same in our experiments because testing considered only the lower confidence limits. This is related to the fact that non-inferiority and equivalence were evaluated using the same simulation data.

The limitation of our experiments is that the number of simulations was necessarily small because of the high cost of implementing the permutation method and randomization method, and N was set to 200 to give a more realistic sample size in clinical studies.

Regarding research question (i), we found that the naïve model (only intercept and treatment group) and model 1 (practical) were conservative, and model 2 (close to data-generating model) was slightly inflated for normal variables. For TTE (RMST) variables, there were some cases of a slightly inflated type-I error. For binominal and TTE (Cox) variables, there was an overall type-I inflation trend, although this was close to the nominal significance level in model 2, making it difficult to evaluate because of the bias in parameter estimation.

For research question (ii), the use of the randomization method is valid for normal variables because it can reproduce the null distribution of the test statistic in CAR, but the permutation method is not valid. TTE (RMST) variables tend to inflate the type-I error when N is small (as described later). Thus, the type-I error rate was inflated when either the permutation or randomization method is used, but when N is sufficient, the use of the permutation method may be considered appropriate, as for normal variables.

In terms of research question (iii), the confidence intervals were reconstructed by replacing the nominal test statistic with each percentage point ($q_p^{\alpha}$ and $q_p^{\alpha-1}$) of the randomly generated test statistic.

If the model can be correctly identified, the type-I error rate can be maintained near the nominal level in non-inferiority and equivalence trials, regardless of the randomization method and variable, and the statistical power is high. Therefore, even without complete randomization, it is not necessary to perform permutated testing and randomization testing as a sensitivity analysis, as described in the FDA guidance. However, if the model may have been misidentified, the following precautions should be taken depending on the analysis model used. In the case of using the normal analysis method, sensitivity analyses using randomization testing may be necessary because the type-I error rate is excessively controlled, resulting in low power. In the case of binominal or TTE (Cox) variables, the most important point is to identify the model correctly, because the type-I error rate is highly inflated even if permutation testing or randomization testing is performed. Because the correct model is unknown, sensitivity analyses using various models would be necessary. If the RMST model is applied to a study with a small sample size, the naïve model under $T_{CAR}$ and $T_{SPBR}$ should be used instead of models with covariates and the approximate true value should be calculated.

## 6. Conclusion

For normal variables, the adjusted confidence interval using the randomization method closes the type-I error rate to the nominal significance level for non-inferiority and equivalence trials with CAR. In addition, for RMST models, the adjusted confidence interval using the randomization method is expected to close the type-I error rate to the nominal significance level by removing sample size bias. The logistic model for binary variables and the Cox model for TTE variables lose consistency when the model is mis-specified, and the adjusted confidence intervals using the randomization method or permutation method cannot control the type-I error rate for the null hypotheses of non-inferiority and equivalence with group differences. Hence, the logistic and Cox models require variable selection and sensitivity analysis.

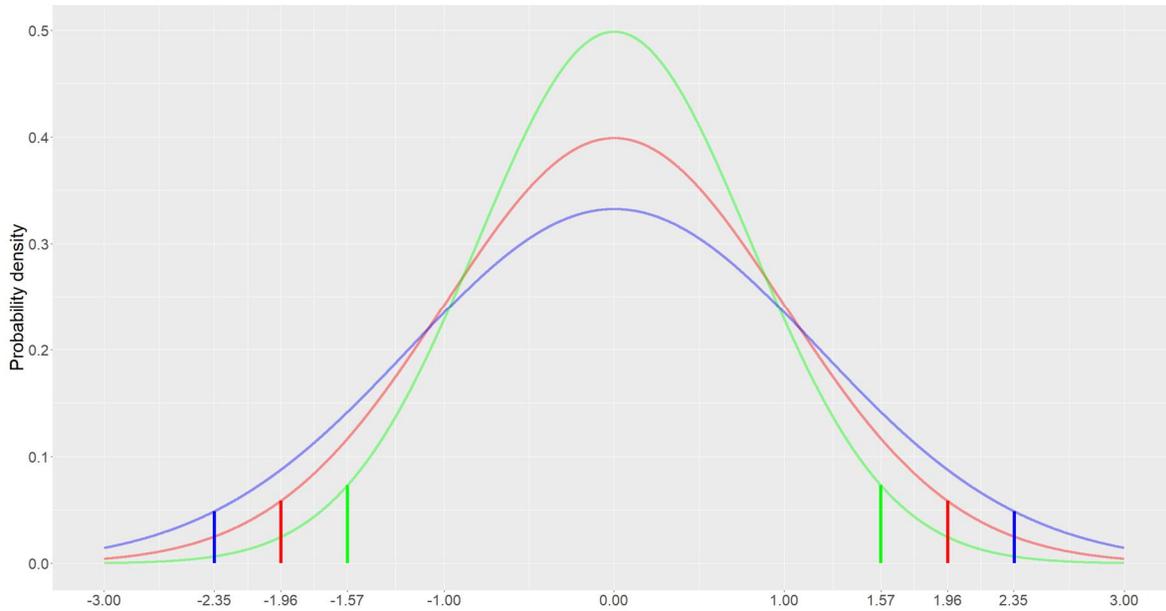

**Figure 1. Change of critical values**

Red line: standard normal distribution, blue line: normal distribution with a larger variance than the standard normal distribution, green line: normal distribution with a smaller variance than the standard normal distribution

**Table 1. Bias of estimator of treatment group**

| Variable | Model | Treatment effect between groups | | | |
|---|---|---|---|---|---|
| | | Presence | | Absence | |
| | | $B_{CAR}$ | $B_{SPBR}$ | $B_{CAR}$ | $B_{SPBR}$ |
| Normal | Naïve | -0.010 | -0.034 | -0.004 | -0.013 |
| | Model 1 | -0.012 | -0.031 | -0.005 | -0.004 |
| | Model 2 | -0.014 | -0.026 | 0.005 | 0.018 |
| Binomial | Naïve | 0.091 | 0.071 | -0.005 | 0.005 |
| | Model 1 | 0.080 | 0.062 | -0.005 | 0.005 |
| | Model 2 | 0.027 | 0.006 | 0.001 | 0.003 |
| TTE (Cox) | Naïve | 0.115 | 0.113 | 0.009 | -0.007 |
| | Model 1 | 0.059 | 0.062 | 0.011 | -0.007 |
| | Model 2 | -0.032 | -0.034 | 0.009 | -0.010 |
| TTE (RMST) | Naïve | 0.008 | -0.003 | 0.002 | -0.008 |
| | Model 1 | 0.009 | -0.004 | 0.002 | -0.010 |
| | Model 2 | 0.009 | -0.001 | 0.001 | -0.008 |

TTE: time-to-event, RMST: restricted mean survival time, CAR: covariate-adaptive randomization, SPBR: stratified permuted block randomization, $B_{CAR}$: average bias (estimated value – true value) of the parameter for the treatment group variable under CAR, $B_{SPBR}$: average bias (estimated value – true value) of the parameter for the treatment group variable under SPBR, Model 1: results given by adjusted model 1, Model 2: results given by adjusted model 2.

**Table 2. Type-I error rate (%)**

| Non-inferiority | | | | | | | |
|---|---|---|---|---|---|---|---|
| Variable | Model | Test Procedure | | | | | |
| | | $T_{CAR}$ | $PT_{CAR}$ | $RT_{CAR}$ | $T_{SPBR}$ | $PT_{SPBR}$ | $RT_{SPBR}$ |
| Normal | Naïve | 0.4 | 0.7 | 1.8 | 0.5 | 0.5 | 2.7 |
| | Model 1 | 1.2 | 0.7 | 2.0 | 0.6 | 0.5 | 3.5 |
| | Model 2 | 3.0 | 2.4 | 2.4 | 3.4 | 2.4 | 2.4 |
| Binomial | Naïve | 4.0 | 5.2 | 3.9 | 2.5 | 4.1 | 4.8 |
| | Model 1 | 3.4 | 4.6 | 3.7 | 2.8 | 3.5 | 4.3 |
| | Model 2 | 2.7 | 3.4 | 2.8 | 1.3 | 2.5 | 3.2 |
| TTE (Cox) | Naïve | 8.3 | 8.7 | 11.2 | 8.5 | 9.4 | 14.2 |
| | Model 1 | 4.9 | 5.4 | 5.7 | 4.4 | 3.8 | 7.2 |
| | Model 2 | 3.2 | 2.3 | 1.5 | 2.6 | 1.5 | 2.5 |
| TTE (RMST) | Naïve | 2.4 | 2.4 | 3.9 | 2.2 | 2.0 | 2.9 |
| | Model 1 | 3.7 | 3.5 | 4.0 | 1.9 | 1.8 | 2.8 |
| | Model 2 | 3.1 | 2.9 | 2.8 | 3.4 | 3.2 | 3.5 |
| Equivalence | | | | | | | |
| Variable | Model | Test Procedure | | | | | |
| | | $T_{CAR}$ | $PT_{CAR}$ | $RT_{CAR}$ | $T_{SPBR}$ | $PT_{SPBR}$ | $RT_{SPBR}$ |
| Normal | Naïve | 0.4 | 0.7 | 1.8 | 0.5 | 0.5 | 2.7 |
| | Model 1 | 1.2 | 0.7 | 2.0 | 0.6 | 0.5 | 3.5 |
| | Model 2 | 3.0 | 2.4 | 2.4 | 3.4 | 2.4 | 2.4 |
| Binomial | Naïve | 4.0 | 5.2 | 3.9 | 2.5 | 4.1 | 4.8 |
| | Model 1 | 3.4 | 4.6 | 3.7 | 2.8 | 3.5 | 4.3 |
| | Model 2 | 2.7 | 3.4 | 2.8 | 1.3 | 2.5 | 3.2 |
| TTE (Cox) | Naïve | 8.3 | 8.7 | 11.2 | 8.5 | 9.4 | 14.2 |
| | Model 1 | 4.9 | 5.4 | 5.7 | 4.4 | 3.8 | 7.2 |
| | Model 2 | 3.2 | 2.2 | 1.5 | 2.6 | 1.5 | 2.5 |
| TTE (RMST) | Naïve | 2.2 | 2.1 | 3.9 | 2.2 | 2.0 | 2.9 |
| | Model 1 | 3.7 | 3.5 | 4.0 | 1.9 | 1.8 | 2.8 |
| | Model 2 | 3.1 | 2.9 | 2.8 | 3.4 | 3.2 | 3.5 |

One-sided nominal significance level of 2.5%, TTE: time-to-event, RMST: restricted mean survival time, CAR: covariate-adaptive randomization, SPBR: stratified permuted block randomization, $T_{CAR}$: testing under CAR, $PT_{CAR}$: permutation testing under CAR, $RT_{CAR}$: randomization testing under CAR, $T_{SPBR}$: testing under SPBR, $PT_{SPBR}$: permutation testing under SPBR, $RT_{SPBR}$: randomization testing under SPBR, Model 1: results given by adjusted model 1, Model 2: results given by adjusted model 2.

**Table 3. Power (%)**

| Non-inferiority | | | | | | | |
|---|---|---|---|---|---|---|---|
| Variable | Model | Test Procedure | | | | | |
| | | $T_{CAR}$ | $PT_{CAR}$ | $RT_{CAR}$ | $T_{SPBR}$ | $PT_{SPBR}$ | $RT_{SPBR}$ |
| Normal | Naïve | 65.9 | 64.4 | 84.4 | 65.4 | 63.7 | 82.4 |
| | Model 1 | 85.2 | 83.2 | 92.9 | 85.3 | 83.1 | 94.7 |
| | Model 2 | 95.9 | 96.0 | 96.1 | 96.4 | 96.2 | 95.6 |
| Binomial | Naïve | 76.2 | 78.5 | 78.3 | 79.4 | 80.2 | 78.2 |
| | Model 1 | 75.5 | 77.7 | 77.7 | 78.1 | 79.4 | 77.3 |
| | Model 2 | 74.4 | 73.9 | 74.4 | 75.7 | 77.0 | 74.9 |
| TTE (Cox) | Naïve | 92.8 | 81.9 | 93.1 | 89.5 | 81.7 | 94.7 |
| | Model 1 | 90.4 | 76.5 | 88.2 | 87.2 | 78.2 | 90.7 |
| | Model 2 | 84.9 | 67.3 | 78.3 | 81.6 | 68.1 | 78.4 |
| TTE (RMST) | Naïve | 83.8 | 82.9 | 88.3 | 83.3 | 83.0 | 88.3 |
| | Model 1 | 88.6 | 87.7 | 90.2 | 88.6 | 87.8 | 91.3 |
| | Model 2 | 91.4 | 90.4 | 90.7 | 91.9 | 90.5 | 91.3 |
| Equivalence | | | | | | | |
| Variable | Model | Test Procedure | | | | | |
| | | $T_{CAR}$ | $PT_{CAR}$ | $RT_{CAR}$ | $T_{SPBR}$ | $PT_{SPBR}$ | $RT_{SPBR}$ |
| Normal | Naïve | 29.7 | 28.7 | 68.4 | 29.9 | 29.2 | 67.4 |
| | Model 1 | 70.1 | 64.9 | 86.7 | 71.2 | 68.7 | 90.5 |
| | Model 2 | 92.1 | 90.8 | 92.2 | 92.5 | 92.9 | 91.7 |
| Binomial | Naïve | 53.8 | 57.0 | 58.7 | 57.4 | 60.6 | 58.3 |
| | Model 1 | 53.5 | 55.2 | 56.8 | 56.3 | 58.6 | 56.9 |
| | Model 2 | 48.6 | 49.4 | 50.2 | 51.5 | 53.5 | 50.1 |
| TTE (Cox) | Naïve | 82.2 | 71.9 | 87.0 | 81.3 | 74.3 | 88.1 |
| | Model 1 | 77.2 | 63.3 | 79.0 | 75.8 | 67.0 | 78.7 |
| | Model 2 | 68.0 | 46.3 | 56.5 | 64.1 | 49.5 | 53.6 |
| TTE (RMST) | Naïve | 67.1 | 66.2 | 75.7 | 69.7 | 69.1 | 79.4 |
| | Model 1 | 75.9 | 74.9 | 80.0 | 80.6 | 79.3 | 85.5 |
| | Model 2 | 82.9 | 80.4 | 81.8 | 86.3 | 84.4 | 85.2 |

TTE: time-to-event**,** RMST: restricted mean survival time**,** CAR: **c**ovariate-adaptive randomization**,** SPBR: stratified permuted block randomization**,** $T_{CAR}$: testing under CAR, $PT_{CAR}$: permutation method under CAR, $RT_{CAR}$: randomization method under CAR, $T_{SPBR}$: testing under SPBR, $PT_{SPBR}$: permutation testing under SPBR, $RT_{SPBR}$: randomization testing under SPBR, Model 1: results given by adjusted model 1, Model 2: results given by adjusted model 2.